\documentclass[aps,twocolumn,epsfig,prb]{revtex4}

\usepackage{amsmath}
\usepackage{graphicx}

\begin{document}

\title{Local density of states at zigzag edge of carbon nanotubes and
graphene} 

\author{K. Sasaki}
\email[Email address: ]{sasaken@flex.phys.tohoku.ac.jp}
\affiliation{Department of Physics, Tohoku University and CREST, JST,
Sendai 980-8578, Japan}

\author{K. Sato}
\affiliation{Department of Physics, Tohoku University and CREST, JST,
Sendai 980-8578, Japan}

\author{J. Jiang}
\affiliation{Department of Physics, North Carolina State University,
Raleigh, NC 27695, U.S.A}

\author{R. Saito}
\affiliation{Department of Physics, Tohoku University and CREST, JST,
Sendai 980-8578, Japan}

\author{S. Onari}
\affiliation{Department of Applied Physics, Nagoya University,
Nagoya 464-8603, Japan}

\author{Y. Tanaka}
\affiliation{Department of Applied Physics, Nagoya University,
Nagoya 464-8603, Japan}

\date{\today}
 
\begin{abstract}
 The electron-phonon matrix element for edge states 
 of carbon nanotubes and graphene at zigzag edges
 is calculated 
 for obtaining renormalized energy dispersion of the edge states.
 Self-energy correction by electron-phonon interaction
 contributes to the energy dispersion of edge states 
 whose energy bandwidth is similar to phonon energy.
 Since the energy-uncertainty of the edge state
 is larger than temperature, we conclude that 
 the single-particle picture of the edge state in not appropriate
 when the electron-phonon interaction is taken into account.
 The longitudinal acoustic phonon mode contributes to 
 the matrix element through the on-site deformation potential
 because the wavefunction of the edge state
 has an amplitude only on one of the two sublattices.
 The on-site deformation potentials 
 for the longitudinal and in-plane tangential optical phonon modes
 are enhanced at the boundary.
 The results of local density of states 
 are compared with the recent experimental data of 
 scanning tunneling spectroscopy.
\end{abstract}

\maketitle

\section{introduction}\label{sec:intro}

The local electronic property near the edge of graphene
depends on the lattice structure of the edge.
For example, 
the zigzag edge induces edge states 
which are $\pi$-electron states localized near the edge 
while the armchair edge does not.~\cite{fujita96} 
The edge states which have a flat energy dispersion,
show a peak in local density of states (LDOS) near the Fermi energy.
The peak structure in LDOS
has been observed at the edge of graphene 
by scanning tunneling microscopy (STM) and spectroscopy
(STS).~\cite{niimi05,niimi06,kobayashi05,kobayashi06}
The LDOS peak is a direct evidence of the edge states
because the peak is not observed at an armchair edge
and the height of the peak decreases with increasing a distance
of the STS tip on graphene plane 
from the zigzag edge.
The data on LDOS are useful to understand 
the energy and life-time of an electron in the edge states.
The life-time of the electron is determined by 
electron-phonon (el-ph) interaction and 
the el-ph interaction is important for almost flat energy dispersion
when the Debye energy ($\hbar \omega_{\rm D} \approx 0.2$ eV) 
is comparable to the energy bandwidth.
Thus the el-ph interaction for edge states affects STS spectra 
and is essential for an analysis of STS.
In this paper, we consider self-energy correction for edge states
induced by the el-ph interaction,
and compare the theoretical results of LDOS with experimental
data.~\cite{niimi05,niimi06,kobayashi05}

A complete flat energy dispersion relation of the edge states
is widely recognized by the theory.~\cite{fujita96}
However, STM/STS~\cite{niimi05,niimi06,kobayashi05,kobayashi06} 
and angle-resolved photo-emission spectroscopy (ARPES)~\cite{sugawara06}
show that the edge states have a small but finite energy dispersion.
Using STM/STS at graphene edge, 
Niimi {\it et al.}~\cite{niimi05,niimi06} 
and Kobayashi {\it et al.}~\cite{kobayashi05,kobayashi06} independently
observed a peak in the LDOS
below the Fermi energy by 20 $\sim$ 30 meV.
The peak position relative to the Fermi point ($E_{\rm F}=0$)
shows that the edge states have a finite bandwidth.
Using ARPES,
Sugawara {\it et al}.,~\cite{sugawara06}
observed the Fermi surface of Kish graphite 
and found a weakly dispersive energy band near the Fermi energy.
In the previous paper, we pointed out 
that next nearest-neighbor (nnn) tight-binding Hamiltonian,
${\cal H}_{\rm nnn}$,
is essential for the bandwidth.~\cite{sasaki06apl}
As shown in Sec.~\ref{sec:ed-ph},
${\cal H}_{\rm nnn}$ lowers the energy dispersion
of the edge states as
$E(k) = \gamma_n (2 \cos ka + 1)$ ($2\pi/3 < ka < 4\pi/3$)
where $\gamma_n$ and $a$
is the hopping integral between nnn sites ($\gamma_n \sim 0.3$ eV)
and a lattice constant of graphite ($a=2.46$ \AA),
and the value of $\gamma_n$ is
calculated on the basis of density-functional theory 
by Porezag {\it et al}.~\cite{porezag95} 
Since $ka=\pi$ state is located at the bottom of the band
and $ka \to 2\pi/3$ (or $4\pi/3$) is located at the top of the band,
the bandwidth of the edge states, $W$, is given by $W=\gamma_n$.
However, the observed energy bandwidth (20 $\sim$ 30 meV)
is much smaller than $\gamma_n$.
The reason why observed bandwidth is smaller than $\gamma_n$
is that the self-energy correction $\Sigma(k)$
renormalizes $E(k)$ as $E(k) + {\rm Re}(\Sigma(k))$.
Because the el-ph interaction makes
the effective mass of the edge states large, 
$W$ generally decreases by taking account of el-ph interaction.
It is also noted that 
$\Gamma(k)\equiv-2{\rm Im}(\Sigma(k))$
represents the energy uncertainty of the edge state.
Since the Fermi-Dirac distribution function
has a width of $k_{\rm B} T$ around the Fermi energy ($E_{\rm F}$),
if $\Gamma(k) > k_{\rm B} T$, then
it is not appropriate to treat the edge state by 
a single-particle picture.

The el-ph interaction is 
calculated by the matrix element of deformation potential.
When the wavefunction is expanded by tight-binding orbitals,
the matrix element consists of
on-site and off-site 
atomic deformation potentials.~\cite{jiang05}
The el-ph matrix element of a given wavefunction
is given by the sum of atomic deformation potentials
over all carbon sites on which the electronic wavefunction
has an amplitude.
The el-ph interaction for edge states shows 
a different behavior from that for extended states.
The unit cell of graphene consists of two sublattices; A and B.
The extended states have a finite densities on both sublattices 
while the edge states have density only on one
sublattice.~\cite{fujita96} 
Suzuura and Ando pointed out for the extended states that
the on-site atomic deformation potentials 
at A-atom and B-atom cancel with each other
in the matrix element for a backward scattering process
because of a phase difference of the wavefunctions
for two sublattices.~\cite{suzuura02}
Thus only the off-site atomic deformation potential,
that is generally weaker than the on-site one,
contributes to the backward scattering.
This is consistent with the fact that
a metallic carbon nanotube (CNT) shows the quantum conductance 
and ballistic character at a low
temperature.~\cite{ando98,bachtold00,park04}  
However, the cancellation of on-site deformation potential
does not occur for edge states, 
since the wavefunction has an amplitude only on a sublattice
for the edge state.
Thus we can expect a relatively strong el-ph interaction 
for the edge states
and a large self-energy correction to the edge states.

The el-ph interaction for the edge states
is relevant to many observations in experiment of CNTs.
Superconductivity in CNTs is an important example.
Takesue {\it et al}.~\cite{takesue06} observed a drop of resistivity 
in multi-walled CNTs and pointed out that
the connection of multi-walled CNTs
to Au-electrode is sensitive
for the occurrence of the resistivity drop.
Since the edge states enhance LDOS near the ends of a CNT,
the el-ph interaction should be sensitive to the properties 
at the interface between the CNT and an electrode.
Furthermore, 
we propose that the large LDOS and a strong el-ph interaction 
favor superconducting instability for the edge
states.~\cite{sasaki06super} 
The self-energy correction is important for an estimation 
of the superconducting transition temperature.
Thus a quantitative discussion 
of the el-ph interaction for edge states
will play a decisive role 
for a future work on STS and superconductivity
of graphene based materials.

This paper is organized as follows.
In Sec.~\ref{sec:ed-ph} 
we give a method for calculating el-ph interaction for edge states.
In Sec.~\ref{sec:result},
we show phonon mode and localization length dependence 
of el-ph interaction. 
In Sec.~\ref{sec:ldos_edge}, 
we calculate self-energy correction to edge states, by which
we obtain a renormalized energy dispersion relation and LDOS.
The theoretical results are compared with the experiment.
Discussions and summary are given in Sec.~\ref{sec:dis}.

\section{Electron-phonon interaction for edge state}\label{sec:ed-ph}

\subsection{Edge States}

\begin{figure}[htbp]
 \begin{center}
  \includegraphics[scale=0.4]{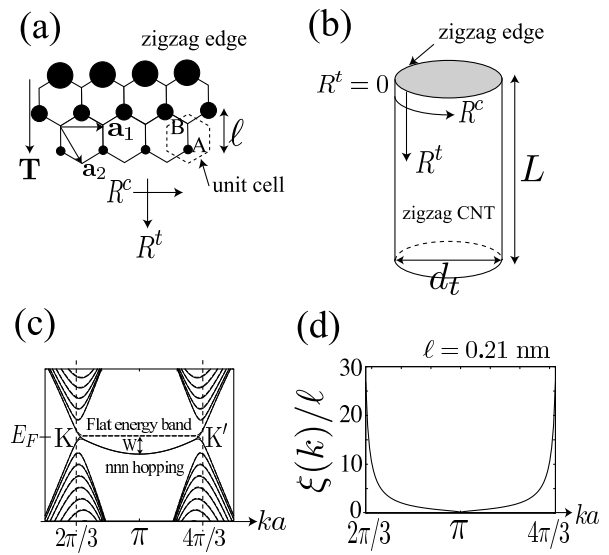}
 \end{center}
 \caption{
 (a) 
 The unit vectors of graphene are 
 denoted by ${\bf a}_1$ and ${\bf a}_2$. 
 ${\bf a}_1$ is the unit vector around the tube axis and
 ${\bf T}$ ($\equiv 2 {\bf a}_2 -{\bf a}_1$) is the 
 translational vector along the zigzag tube axis.
 We define $2\ell \equiv |{\bf T}|$ where $\ell = 0.21$ nm
 and $a \equiv |{\bf a}_1|$.
 The density of an edge state has a value only on A-atoms,
 which is represented by solid circles. 
 The radius of a solid circle is proportional to the density,
 which shows the localization.
 (b) The edge states exist near the zigzag edge
 ($R^t=0$) of the $(n,0)$ CNT with open boundary.
 The diameter and length of the CNT is denoted by $d_t$ and $L$, respectively.
 (c) ${\cal H}_{\rm nn}$ has a flat energy band 
 of the edge states 
 between the K and K' points at the Fermi energy ($E_{\rm F}=0$).
 ${\cal H}_{\rm nnn}$ causes a bandwidth $W=\gamma_n$.
 (d) We plot the localization length,
 $\xi(k)/\ell=-1/\ln(2|\cos(ka/2)|)$ for $2\pi/3 < ka < 4\pi/3$.
 }
 \label{fig:edgestate}
\end{figure}

The edge states of graphene 
are $\pi$-electronic states
localized near the zigzag edge
(Fig.~\ref{fig:edgestate}(a)).~\cite{fujita96}
Since a CNT is a rolled-up sheet of graphene,
the edge states may exist 
near the open boundaries of $(n,0)$ zigzag CNT 
regardless of the value of $n$
(see Fig.~\ref{fig:edgestate}(b)).
The energy eigen equation of the
nearest-neighbor (nn) tight-binding Hamiltonian,
${\cal H}_{\rm nn}|\Psi_k\rangle = E(k)|\Psi_k\rangle$,
gives a flat energy band, $E(k)=0$, for 
the edge states where $k$ 
is the wavevector along the edge.
As shown in Fig.~\ref{fig:edgestate}(c),
the edge state exists when $2\pi/3 < ka < 4\pi/3$
and ${\cal H}_{\rm nnn}$ lowers the energy dispersion
of the edge states as
$E(k) = W (2 \cos ka + 1)$ ($2\pi/3 < ka < 4\pi/3$).~\cite{sasaki06apl}
The edge state behaves as a plane wave around the axis
($R^c$ in Fig.~\ref{fig:edgestate}(b)), while
wavefunction is localized in the direction of nanotube axis $R^t$.
The localization length
in the direction perpendicular to the edge is $k$-dependent, 
$\xi(k)=-|{\bf T}|/2\ln(2|\cos(ka/2)|)$
(see Fig.~\ref{fig:edgestate}(d)).~\cite{sasaki05prb}
Though $\xi(k)$ becomes infinite
when $ka \to 2\pi/3$ or $ka \to 4\pi/3$ for a graphene,
we can show that $\xi(k) \alt d_t/2$ for a CNT
where $d_t \equiv na/\pi$ is the diameter of the zigzag CNT. 
To explain this, 
let us consider 
a metallic zigzag CNT, $(3m,0)$.
Then $k$ for the edge states are discrete 
due to the periodic boundary condition,
which is given by
$k(i) = 2\pi/3a + 2\pi i/3ma$
($i=1,\ldots,m-1$).
We get the largest localization length 
$\xi(k(1)) = \xi(k(m-1)) \approx d_t/2$.

The wavefunction of the edge state 
is written as~\cite{sasaki06apl}
\begin{align}
 | \Psi_k \rangle 
 = \frac{N_k}{\sqrt{n}}
 \sum_{u} \exp \left\{ ik R^c_{u,A}
 -\frac{R^t_{u,A}}{\xi(k)} \right\}
 | \phi({\bf R}_{u,A}) \rangle,
 \label{eq:wf_edge}
\end{align}
where $| \phi({\bf R}_{u,s}) \rangle$
is 2${\rm p}_z$ orbital of a carbon atom and
$N_k$ is a normalization constant.
The summation on $u$
is taken for all unit cells of graphene or a CNT.
We take the coordinate ${\bf R}=(R^c,R^t)$ on the cylindrical surface of
a zigzag CNT, in which $R^c$ and $R^t$ are coordinates around and along
the tube axis, respectively (Fig.~\ref{fig:edgestate}(b)).  
The position of a carbon atom 
is denoted by ${\bf R}_p \equiv (R^c_{u,s},R^t_{u,s})$
where $p=(u,s)$ represents the $s$-th sublattice ($s=A,B$) 
in the $u$-th hexagonal unit cell.
As is taken for the zigzag edge site $R^t_{u,A}=0$,
the edge state has amplitudes only on A-atoms ($s=A$).
Equation~(\ref{eq:wf_edge}) is correct for $2\pi/3 < ka \le \pi$.
For $\pi < ka < 4\pi/3$, 
a phase factor $\exp\{i\pi R^t_{u,A}/\ell\}$ 
should be multiplied to
$| \phi({\bf R}_{u,A}) \rangle$.~\cite{sasaki05prb}

\subsection{Electron-Phonon Interaction}

The el-ph interaction for graphene is formulated by 
Jiang {\it et al}.~\cite{jiang05}, 
which will be applied to the edge states.
The el-ph interaction for the edge states is written as
\begin{align}
 {\cal H}_{\rm int} =
 \frac{1}{\sqrt{N_{\rm u}}} \sum_{k,k'}
 \sum_{q_t,\nu}
 \alpha^\nu_{k k'}({\bf q}) 
 (b_{{\bf q},\nu}+b_{{\bf -q},\nu}^\dagger)
 c_{k'}^\dagger c_k,
 \label{eq:Hint}
\end{align}
where $N_{\rm u}$ is the number of graphite unit cells,
$c_k$ is the annihilation operator of the edge state,
and $b_{{\bf q},\nu}$ is the annihilation operator
of the $\nu$-th phonon mode.
There are six phonon modes:
out-of-plane tangential acoustic/optical mode (oTA/oTO),
in-plane tangential acoustic/optical mode (iTA/iTO), and
longitudinal acoustic/optical mode (LA/LO).~\cite{saito98book}
$\alpha^\nu_{kk'}({\bf q})$
is the el-ph interaction connecting two edge
states $k$ and $k'$ by $\nu$-th phonon mode with momentum ${\bf q}$.
Due to the momentum conservation along the edge,
$k'=k + q$,
while the wavevector perpendicular to the edge
$q_t$ 
is needed to sum over the Brillouin zone.
$\alpha^\nu_{kk'}({\bf q})$ is given by 
$\alpha^\nu_{kk'}({\bf q})
\equiv A^\nu({\bf q}) M^\nu_{kk'}({\bf q})/\sqrt{2}$ where
$A^\nu({\bf q})=\sqrt{\hbar/m_c \omega_\nu({\bf q})}$
is the amplitude of phonon
($\hbar \omega_\nu({\bf q})$ is the energy of the 
$\nu$-th phonon with the momentum
${\bf q}$) and 
$M^\nu_{kk'}({\bf q})$ is the el-ph matrix element,
\begin{align}
 M^\nu_{kk'}({\bf q}) \equiv  -\sum_{p} \langle \Psi_{k'} |
 \nabla v({\bf R}_p) | \Psi_k \rangle
 \cdot U({\bf R}_p){\bf e}^\nu_{\bf q}(s)
 e^{i{\bf q}\cdot {\bf R}_p}.
 \label{eq:M}
\end{align}
Here $v({\bf R}_p)$ is the Kohn-Sham potential 
of a neutral pseudoatom calculated on the basis of density-functional
theory by Porezag {\it et al}.~\cite{porezag95}
for a carbon atom at ${\bf R}_p$, 
${\bf e}^\nu_{\bf q}(s)$ is phonon eigenvector
at an $s$-atom normalized in the unit cell as
$\sum_{s=A,B} {\bf e}^\nu_{\bf q}(s)^* \cdot {\bf e}^\nu_{\bf q}(s) =1$,
and $U({\bf R}_p)$ is a rotational operator 
for ${\bf e}^\nu_{\bf q}(s)$
from an $s$-th atom at origin
to a $s$-th atom at ${\bf R}_p$.
To obtain $\omega_\nu({\bf q})$ and ${\bf e}^\nu_{\bf q}(s)$,
we use the force constant parameters calculated by
Dubay and Kresse~\cite{dubay03}
for the dynamical matrix.~\cite{saito98book}

Putting Eq.~(\ref{eq:wf_edge}) into Eq.~(\ref{eq:M}),
we obtain
\begin{align}
 & M^\nu_{k k'}({\bf q})
 = -\frac{N_{k'}N_{k}}{n} 
 \notag \\
 & \times \sum_{u',u}
 \exp \left\{ -ik' R^c_{u',A} + ik R^c_{u,A}
 -\frac{R^t_{u',A}}{\xi(k')} 
 -\frac{R^t_{u,A}}{\xi(k)} \right\}
 \notag \\
 & \times m({\bf R}_{u',A},{\bf R}_{u,A};{\bf q},\nu),
 \label{eq:m^nu}
\end{align}
where $m({\bf R}_{u',A},{\bf R}_{u,A};{\bf q},\nu)$ 
is the atomic deformation potential,~\cite{jiang05} defined by
\begin{align}
 & m({\bf R}_{u',A},{\bf R}_{u,A};{\bf q},\nu) 
 \notag \\
 & \equiv \sum_p
 \langle \phi({\bf R}_{u',A}) |
 \nabla v({\bf R}_p)
 | \phi({\bf R}_{u,A}) \rangle
 \cdot U({\bf R}_p){\bf e}^\nu_{\bf q}(s)
 e^{i{\bf q} \cdot {\bf R}_p}.
 \label{eq:adp}
\end{align}
There are two types of 
$m({\bf R}_{u',s'},{\bf R}_{u,s};{\bf q},\nu)$.
The first type is the case of $(u',s')=(u,s)$ 
which is referred to as 
the {\it on-site} atomic deformation potential.~\cite{jiang05}
The other one is $(u',s')\ne (u,s)$ which is 
the {\it off-site} atomic deformation potential.
The on-site (off-site) atomic deformation potential 
represents a scattering process of an electron 
from ${\bf R}_{u,s}$ to the same (different) site.

The off-site deformation potential matrix element 
for the next-nearest A atoms,
$|\langle \phi({\bf R}_{u',A})|\nabla v({\bf R}_p)|
\phi({\bf R}_{u,A})\rangle \cdot {\hat n}|$, 
is negligible (for any ${\bf R}_p$) in Eq.~(\ref{eq:adp})
because it decay quickly as a function of 
$|{\bf R}_{u',A}-{\bf R}_{u',A}|$ where 
${\hat n}$ is a unit vector along the two carbon
atoms.
Density-functional theory gives that
the off-site deformation potential matrix element 
for nearest-neighbor interaction is 
$|\langle \phi({\bf R}_{u,B})|\nabla v({\bf R}_{u,B})|
\phi({\bf R}_{u,A})\rangle \cdot {\hat n}|\sim $ 
3eV/\AA.~\cite{jiang05,porezag95} 
As we pointed out above,
the wavefunction of the edge state
has an amplitude only of one sublattice and thus
this nearest-neighbor term does not contribute to
$\alpha^\nu_{kk'}({\bf q})$.
Thus the off-site atomic deformation potential
does not contribute to $\alpha^\nu_{kk'}({\bf q})$
for the edge states.
For the on-site atomic deformation potential,
we need to consider several carbon atoms 
which are located near ${\bf R}_{u,A}$ 
for the center of deformation potential 
${\bf R}_p$ in Eq.~(\ref{eq:adp}).
The value of 
$|\langle \phi({\bf R}_{u,A})|\nabla v({\bf R}_p)|
\phi({\bf R}_{u,A})\rangle \cdot {\hat n}|$
is not negligible if 
$|{\bf R}_p-{\bf R}_{u,A}| \le 3$\AA.
Density-functional theory gives that
the largest contribution from nearest-neighbor site is 
$|\langle \phi({\bf R}_{u,A})|\nabla v({\bf R}_{u,B})|
\phi({\bf R}_{u,A})\rangle \cdot {\hat n}| \sim 8$
eV/\AA.~\cite{porezag95} 
Thus on-site deformation potential is more important than 
the off-site deformation potential.~\cite{suzuura02}

Now we can write
$m({\bf R}_{u',A},{\bf R}_{u,A};{\bf q},\nu)$ as
\begin{align}
 m({\bf R}_{u',A},{\bf R}_{u,A};{\bf q},\nu)
 = \delta_{u'u} 
 m_{\rm on}(R_{u,A}^t;{\bf q},\nu)
 e^{i{\bf q} \cdot {\bf R}_{u,A}},
 \label{eq:m_on_def}
\end{align}
where $m_{\rm on}(R_{u,A}^t;{\bf q},\nu)$ is
on-site deformation potential from all possible ${\bf R}_p$,
defined as
\begin{align}
 & m_{\rm on}(R_{u,A}^t;{\bf q},\nu) \equiv \notag \\
 & \sum_p
 \langle \phi({\bf R}_{u,A}) |
  \nabla v({\bf R}_p) 
 | \phi({\bf R}_{u,A}) \rangle
 \cdot 
 U({\bf R}_p) {\bf e}^\nu_{\bf q}(s)
 e^{i{\bf q} \cdot ({\bf R}_p-{\bf R}_{u,A})}.
 \label{eq:mon}
\end{align}
Putting Eq.~(\ref{eq:m_on_def}) into Eq.~(\ref{eq:m^nu}),
we get 
\begin{align}
 &M^\nu_{k k'}({\bf q})
 =-N_{k'}N_{k} \delta_{k',k+q}
 \notag \\
 & \sum_{R_{u,A}^t}
 \exp\left\{ \left( i q_t
 -\frac{1}{\xi(k')} -\frac{1}{\xi(k)} \right) R^t_{u,A}
 \right\}
 m_{\rm on}(R_{u,A}^t;{\bf q},\nu).
 \label{eq:m_formula}
\end{align}
For $ka \le \pi < k'a$ or $k'a \le \pi < ka$,
we must multiply 
$\exp\{i\pi R^t_{u,A}/\ell\}$ to 
$m_{\rm on}(R_{u,A}^t;{\bf q},\nu)$ in 
Eq.~(\ref{eq:m_formula}).

It is noted that $m_{\rm on}(R_{u,A}^t;{\bf q},\nu)$ 
is defined to be independent of $R^c_{u,A}$.
Instead, $R^c_{u,A}$ appears in the phase of 
$m({\bf R}_{u,A},{\bf R}_{u,A};{\bf q},\nu)$
in Eq.~(\ref{eq:m_on_def}).
The phase gives the momentum conservation 
around the tube axis, $\delta_{k',k+q}$,
after the summation about $R_{u,A}^c$ is made 
in Eq.~(\ref{eq:m^nu}).
On the other hand,
$m_{\rm on}(R_{u,A}^t;{\bf q},\nu)$ depends on $R_{u,A}^t$
because $\sum_p$ in Eq.~(\ref{eq:mon}) is restricted for
$R^t_p \ge 0$ and $m_{\rm on}(R_{u,A}^t;{\bf q},\nu)$
does not have a translational symmetry near the edge.
Hereafter, we refer $m_{\rm on}(R_{u,A}^t;{\bf q},\nu)$
as boundary deformation potential.

$m_{\rm on}(R_{u,A}^t;{\bf q},\nu)$
depends on $\nu$ strongly.
The iTO and LO modes whose eigenvectors are pointing 
in the direction perpendicular to the edge 
give a large contribution to $m_{\rm on}(R_{u,A}^t;{\bf q},\nu)$.
We generally expect that
the boundary deformation potential appears only near the edge.
In fact,
$|\langle \phi({\bf R}_{u,A})|\nabla v({\bf R}_p)|\phi({\bf R}_{u,A})
\rangle \cdot {\hat n}|$
is finite only when 
$|{\bf R}_p-{\bf R}_{u,A}| \le 3$\AA.
Thus $m_{\rm on}(R_{u,A}^t;{\bf q},\nu)$
depends on $R^t_{u,A}$ only when $R^t_{u,A} \le 3$\AA.
The boundary deformation potential 
contributes to the matrix element increasingly
with decreasing $\xi(k)$ or $\xi(k')$,
which can be seen in Eq.~(\ref{eq:m_formula}).

\begin{figure}[htbp]
 \begin{center}
  \includegraphics[scale=0.43]{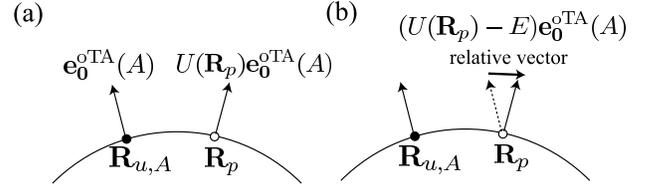}
 \end{center}
 \caption{(a) For the oTA mode with ${\bf q}={\bf 0}$, 
 the eigenvectors of two atoms 
 at ${\bf R}_{u,A}$ and ${\bf R}_p$ are 
 pointing perpendicular to the CNT surface.
 (b) In order to calculate the atomic deformation potential,
 one needs to consider the relative displacement 
 of two carbon atoms.
 }
 \label{fig:relative}
\end{figure}

In Eq.~(\ref{eq:mon}), 
we must consider relative atomic motion 
of ${\bf R}_{u,A}$ to ${\bf R}_p$.
This can be done by replacing 
$U({\bf R}_p)$ with $U({\bf R}_p)-E$ where
$E$ is the unit matrix.
In Fig.~\ref{fig:relative}(a)
we show phonon eigenvector of the oTA mode
pointing perpendicular to the nanotube surface.
If we consider the phonon eigenvector in a flat surface or graphene,
the relative displacement of the two atoms
becomes zero when ${\bf q}={\bf 0}$.
However, it is not the case for a cylindrical surface.
As shown in Fig.~\ref{fig:relative}(b),
the relative eigenvector, 
$(U({\bf R}_p)-E){\bf e}^{\rm oTA}_{\bf q}(s)$,
remains finite.
The relative vector for the oTA mode changes the area 
on a cylindrical surface
and is similar to the LA mode 
except for a $1/d_t$ reduction of the amplitude, by which
$m_{\rm on}(R_{u,A}^t;{\bf q},{\rm oTA})$ 
is proportional to $1/d_t$.
On the other hand,
the relative displacement becomes zero for 
the in-plane modes such as iTA and LA, 
when ${\bf q} \to {\bf 0}$.
Thus, the el-ph interaction by the iTA and LA modes
vanish for ${\bf q} = {\bf 0}$ while
the oTA mode provides a finite 
$m_{\rm on}(R_{u,A}^t;{\bf q},{\rm oTA})$
even for ${\bf q} = {\bf 0}$.
Hereafter, we simply use 
$U({\bf R}_p)$ for $U({\bf R}_p)-E$.

\section{Calculated Results}\label{sec:result}

In this section,
we plot $|M_{kk'}^\nu({\bf q})|$ for several values of
$k$ and $k'$,
and examine the dependence of $|M_{kk'}^\nu({\bf q})|$
on phonon mode $\nu$ in Sec.~\ref{subsec:M_pm} and 
on nanotube diameter $d_t$ in Sec.~\ref{subsec:dt}.

\subsubsection{$\nu$ dependence of $|M_{kk'}^\nu({\bf
   q})|$}\label{subsec:M_pm}

\begin{figure*}[htbp]
 \begin{center}
  \includegraphics[scale=0.85]{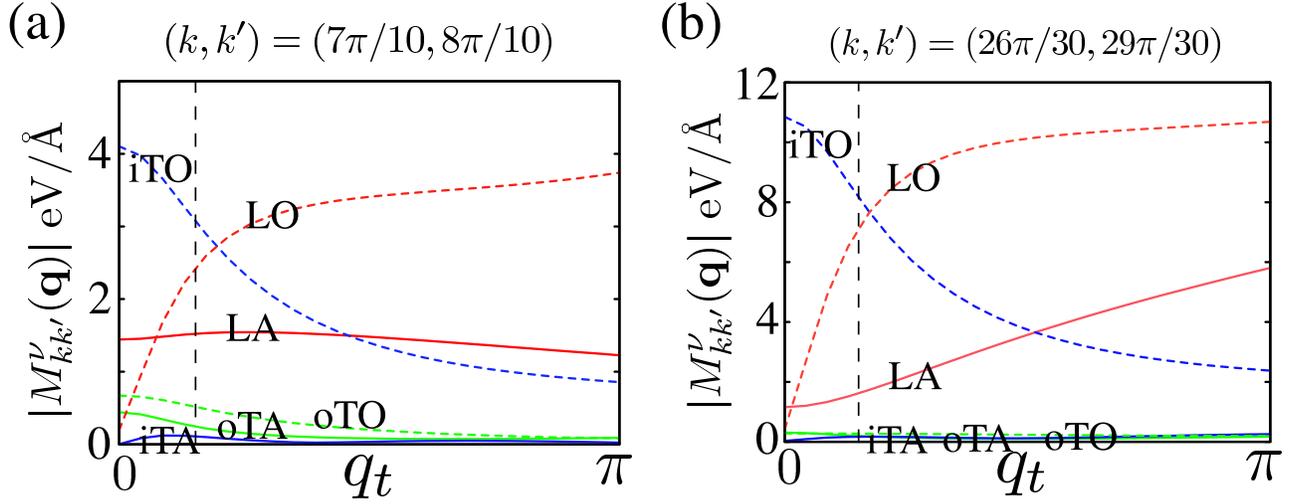}
 \end{center}
 \caption{(Color online)
 $|M_{kk'}^\nu({\bf q})|$ for the $(60,0)$ zigzag CNT:
 (a) $(k,k')=(7\pi/10,8\pi/10)$ and 
 (b) $(k,k')=(26\pi/30,29\pi/30)$
 are plotted as a function of $q_t$ where $q=\pi/10$.
 Three solid curves represent acoustic phonon modes:
 oTA(green), iTA(blue) and LA (red),
 and three dashed curves are optical modes:
 oTO(green), iTO(blue) and LO (red).
 The vertical dashed lines represent $q_t=\sqrt{3} q$ 
 ($\sqrt{3} = |{\bf T}|/|{\bf a}_1|$).
 }
 \label{fig:Me_14}
\end{figure*}

In the calculation,
we consider the magnitude of $M_{kk'}^\nu({\bf q})$
for the $(60,0)$ CNT ($d_t \approx 5$ nm).
In Fig.~\ref{fig:Me_14},
we plot $|M_{kk'}^\nu({\bf q})|$ as a function of $q_t$
for (a) $(k,k')=(7\pi/10,8\pi/10)$ and (b) $(26\pi/30,29\pi/30)$. 
The corresponding phonon eigenvector is the same ($q=\pi/10$).
Thus the difference between the two cases
shows the dependence of $|M_{kk'}^\nu({\bf q})|$
on $\xi(k)$ and $\xi(k')$.
$(7\pi/10,8\pi/10)$ is chosen as an example that
$\xi(k)$ ($=22$\AA)
and $\xi(k')$ ($=4.4$\AA) are longer than 
the carbon-carbon bond length ($a_{\rm cc}=1.4$\AA), while
$(26\pi/30,29\pi/30)$ is chosen as an example that
$\xi(k)$ ($= 2.4$\AA)
and $\xi(k')$ ($= 0.9$\AA) are comparable to $a_{\rm cc}$.

As shown in Fig.~\ref{fig:Me_14}(a),
$|M_{kk'}^{\rm iTO}({\bf q})|$
is 4 eV/\AA \ at $q_t=0$.
$|M_{kk'}^{\rm iTO}({\bf q})|$
decreases with increasing $q_t$,
while $|M_{kk'}^{\rm LO}({\bf q})|$ 
increases with increasing $q_t$.
This behavior of the iTO and LO modes
relates to the boundary deformation potential.
The eigenvector of the iTO (LO) mode 
is pointing along the tube axis 
when $q_t/|{\bf T}| < q/|{\bf a}_1|$ 
($q_t/|{\bf T}| > q/|{\bf a}_1|$)
and then produces a large boundary deformation potential.
On the other hand, ${\bf e}^{\rm oTO}_{\bf q}(s)$
is perpendicular to the CNT axis
and the oTO mode does not contribute to a boundary deformation
potential.
The value of $|M_{kk'}^{\rm oTO}({\bf q})|$
is considerably smaller than $|M_{kk'}^{\nu}({\bf q})|$ 
for the iTO and LO modes.
Thus the contribution of the oTO mode to the el-ph interaction 
can be neglected for $d_t \approx 5$ nm CNT.
In Fig.~\ref{fig:Me_14}(b), 
$|M_{kk'}^{\rm iTO}({\bf q})|$ and $|M_{kk'}^{\rm LO}({\bf q})|$
reach 11 eV/\AA, which indicates that
$|M_{kk'}^{\nu}({\bf q})|$ for the iTO and LO modes increase
significantly with decreasing $\xi(k)$ and $\xi(k')$.
The boundary deformation potential becomes more effective 
with decreasing the localization length.

The matrix element for iTA mode is smaller than 
the other acoustic modes for a wide range of $q_t$
as shown in Fig.~\ref{fig:Me_14}(a).
Though $|M_{kk'}^{\rm iTA}({\bf q})|$ can be comparable to 
$|M_{kk'}^{\rm oTA}({\bf q})|$ as shown in Fig.~\ref{fig:Me_14}(b),
the contribution of the iTA mode to the el-ph interaction
is negligible to that for oTA
because the amplitude of the iTA mode 
is smaller than that of the oTA mode;
$A^{\rm iTA}({\bf q}) \alt A^{\rm oTA}({\bf q})$.
On the other hand, 
the oTA and LA modes are important for lower temperature
in the el-ph interaction.
The oTA mode changes the volume of a CNT
and gives an on-site deformation potential
as shown in Fig.~\ref{fig:relative}.
Moreover,
the energy of the oTA mode is smallest among acoustic modes
and thus $A^{\rm oTA}({\bf q})$ can be larger
than $A^{\rm LA}({\bf q})$.
The LA mode is a area-changing mode
and produces a large on-site deformation potential, 
and contributes to the matrix element 
most significantly among acoustic modes.

\subsubsection{$d_t$ dependence of $|M_{kk'}^\nu({\bf
   q})|$}\label{subsec:dt}

\begin{figure}[htbp]
 \begin{center}
  \includegraphics[scale=0.6]{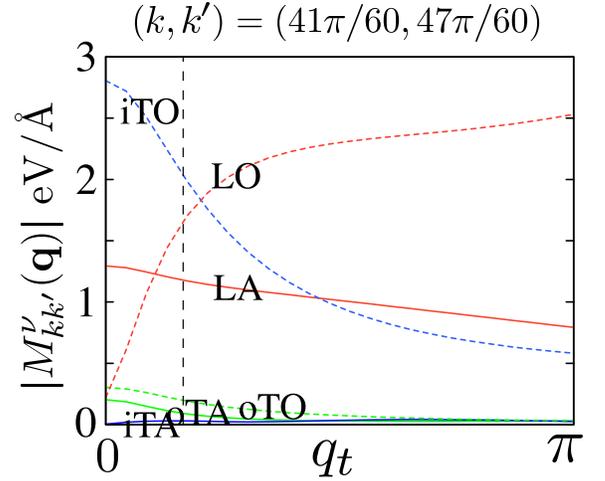}
 \end{center}
 \caption{(Color online)
 $|M_{kk'}^\nu({\bf q})|$ for $(120,0)$ zigzag CNT.
 The phonon eigenvector is the same
 as Fig.~\ref{fig:Me_14}.
 The vertical dashed line represents $q_t=\sqrt{3} q$.
 }
 \label{fig:Me_17}
\end{figure}

It is shown from Eq.~(\ref{eq:m_formula}) that 
$M_{kk'}^\nu({\bf q})$ for two CNTs with 
different diameters ($d_t$) are the same
for the same values of $k$ and $k'$.
In general,
we can find the similar values of $k$ (or $k'$) for different $d_t$.
However, it is not the case for edge states near the K or K' point.
Then $M_{kk'}^\nu({\bf q})$ depends on $d_t$
for such edge states.
For instance, 
$M_{kk'}^\nu({\bf q})$ depends on $d_t$
for the largest $\xi(k) \approx d_t/2$.
In Fig.~\ref{fig:Me_17}, 
we plot $|M_{kk'}^\nu({\bf q})|$ 
for $(41\pi/60,47\pi/60)$ of the $(120,0)$ CNT ($d_t \approx 10$ nm).
$|M_{kk'}^\nu({\bf q})|$
has the similar functional shape as shown in 
Fig.~\ref{fig:Me_14}(a)
while the values become smaller.

The reduction of $|M_{kk'}^\nu({\bf q})|$
can be explained as follows.
If we neglect $d_t$ dependence of the boundary deformation potential,
the summation on $R^t_{u,A}$
in Eq.~(\ref{eq:m_formula}) can be performed analytically.
$M_{kk'}^\nu(q_t=0)$ is then proportional to 
the factor $R_{kk'}(d_t)$;
\begin{align}
 R_{kk'}(d_t) \equiv \frac{\sqrt{\xi(k)\xi(k')}}{\xi(k)+\xi(k')},
 \label{eq:ratioij}
\end{align}
where $R_{kk'}(d_t)$ is a function of $d_t$
because $\xi$ depends on $d_t$.
Since $R_{kk'}(5 {\rm nm})=0.37$ and $R_{kk'}(10 {\rm nm})=0.30$,
we have $R_{kk'}(10 {\rm nm})/R_{kk'}(5 {\rm nm}) \approx 0.81$.
This ratio reproduces 
$|M_{kk'}^{\rm LA}({\bf q})|_{10 {\rm nm}}
/|M_{kk'}^{\rm LA}({\bf q})|_{5 {\rm nm}} = 1.3/1.5 \approx 0.86$ 
at $q_t=0$.

\section{Local density of states}\label{sec:ldos_edge}

Now we calculate 
a renormalized energy dispersion relation and 
LDOS using self-energy $\Sigma(k,i\omega_n)$ 
which is defined by 
\begin{align}
 \Sigma(k,i\omega_n)
 &= \frac{2k_{\rm B} T}{N_{\rm u}} 
 \sum_m \sum_{{\bf q},\nu}
 \frac{|\alpha_{kk+q}^\nu({\bf q})|^2\hbar \omega_\nu({\bf
 q})}{(\omega_n-\omega_m)^2 + (\hbar \omega_\nu({\bf q}))^2} \notag \\
 & \times 
 \frac{1}{i\omega_m - (E(k+q)-E_{\rm F}) - \Sigma(k+q,i\omega_m) },
 \label{eq:selfene}
\end{align}
where $T$ is temperature and 
$\omega_n=\pi k_{\rm B}T (2n+1)$ 
is the Matsubara frequency ($n$ is integer).
We put the cut-off Matsubara frequency,
$|\omega_n| \le \hbar \omega_{\rm D}\approx 0.2$ eV.
By means of Pad\'e approximation,~\cite{vidberg77}
$\Sigma(k,i\omega_n)$ is changed to
that on the real-axis of $\omega$: $\Sigma(k,\omega)$.
Then we find a solution $\epsilon$ satisfying
$\epsilon = (E(k)-E_{\rm F})+{\rm Re}\left( \Sigma(k,\epsilon) \right)$ 
as a function of $k$, $E_{\rm F}$ and $T$.
The obtained $\epsilon_{E_{\rm F},T}(k)$ is 
a renormalized energy dispersion relation.
We adopt $T=50$ K in the following calculations.
The $T$-dependence of $\epsilon_{E_{\rm F},T}(k)$ is negligible for 
a wide range of $T$, for instance, $T=77$ K gives 
almost the same result.
Calculations at low temperature have some numerical advantage 
since the number of $\omega_n$ increases.
The LDOS curve is defined as a function of bias voltage ($V$) 
between graphene and the STS tip, and 
the distance ($R^t$) from the zigzag edge sites and the STS tip by 
\begin{align}
 D(V,R^t)=\frac{1}{\pi}
 \sum_k 
 \frac{\Gamma_{E_{\rm F}}(k)}{(V-\epsilon_{E_{\rm F}}(k))^2 +
 \Gamma_{E_{\rm F}}(k)^2}|\Psi_k(R^t)|^2,
\end{align}
where $\Gamma_{E_{\rm F}}(k)\equiv
-2{\rm Im}(\Sigma(k,\epsilon_{E_{\rm F}}(k)))$
is the width of STS spectrum.
It is noted that the value of $E_{\rm F}$ is given
in such a way that
the number of valence (edge) states is conserved
when we calculate $\Sigma(k,i\omega_n)$ self-consistently,
that is, $\sum_{E(k)< E_{\rm F}} 1 = 
\sum_{\epsilon_{E_{\rm F}}(k) < E_{\rm F}} 1$.

\begin{figure*}[htbp]
 \begin{center}
  \includegraphics[scale=0.7]{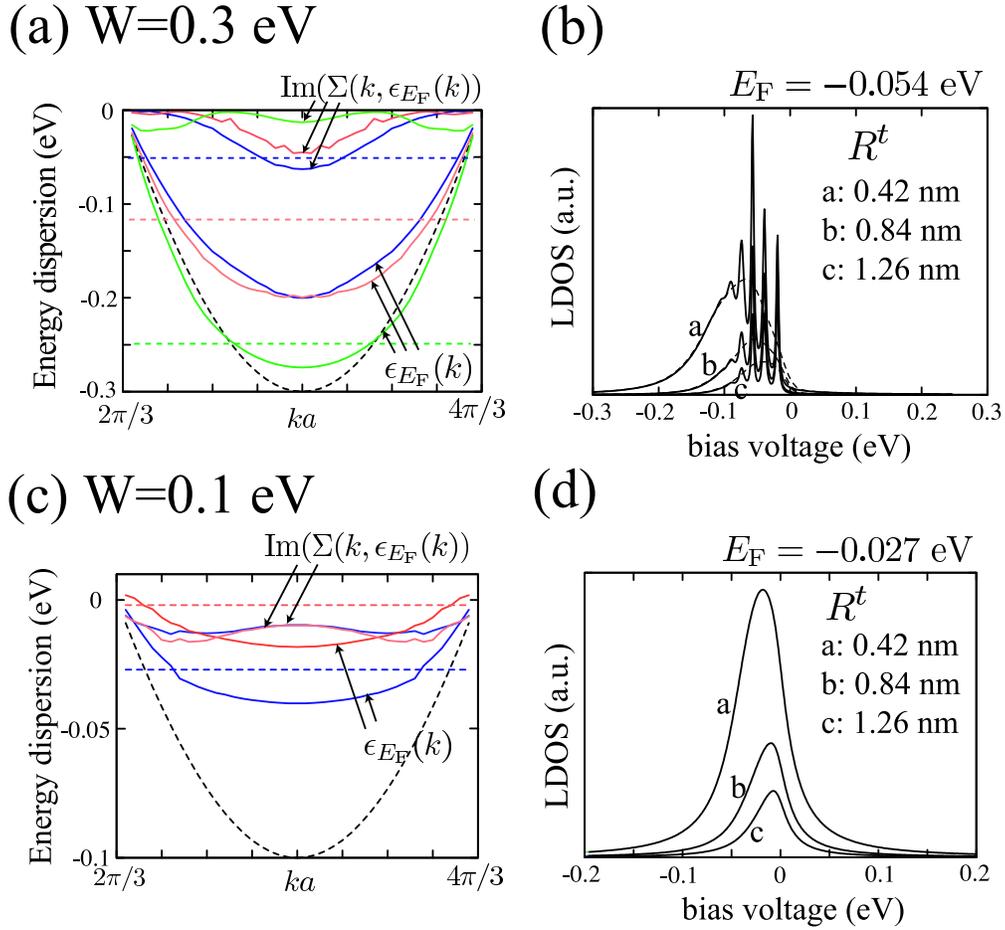}
 \end{center}
 \caption{(Color online)
 Calculated energy dispersion $\epsilon_{E_{\rm F}}(k)$ and 
 LDOS $D(V,R^t)$ for (a,b) $W=0.3$ eV and (c,d) $W=0.1$ eV.
 In (a), $E(k)$ is plotted as the dashed curve, 
 and $\epsilon_{E_{\rm F}}(k)$ is plotted for 
 $E_{\rm F}=-0.054$, $-0.122$, $-0.250$ eV.
 We plot ${\rm Im}(\Sigma(k,\epsilon_{E_{\rm F}}(k)))$ 
 to show the $k$-dependence.
 In (b), $D(V,R^t)$ for $E_{\rm F}=-0.054$ eV
 and its smoothed curve (dashed curve) 
 are plotted for $R^t =0.42$, $0.84$ and $1.26$ nm.
 In (c), $\epsilon_{E_{\rm F}}(k)$ is plotted for 
 $E_{\rm F}=-0.002$ eV and $-0.027$ eV.
 In (d), we plot $D(V,R^t)$ for $E_{\rm F}=-0.027$ eV.
 }
 \label{fig:selfene}
\end{figure*}

\subsection{Numerical Result}

In Fig.~\ref{fig:selfene}(a),
we plot $E(k)$ ($W=0.3$ eV) 
without self-energy correction as the dashed curve
and renormalized energy dispersion $\epsilon_{E_{\rm F}}(k)$ for 
$E_{\rm F}=-0.054$, $-0.122$, $-0.250$ eV
as the blue, red and green curves, respectively.
The reason why we consider different $E_{\rm F}$ values 
is that in the experiment charge transfer from STS tip or substrate 
may modify the $E_{\rm F}$ values
and that we need to investigate $E_{\rm F}$ dependence.
The parallel dashed lines denote the Fermi level for
these $E_{\rm F}$ values.
For each value of $E_{\rm F}$, 
${\rm Im}(\Sigma(k,\epsilon_{E_{\rm F}}(k)))$ is plotted
to show the $k$-dependence.
By denoting the renormalized bandwidth as $W'$,
we obtain $W' \approx 0.2$ eV 
for $E_{\rm F}=-0.054$ eV and $-0.122$ eV.
The corresponding mass enhancement parameter $\lambda$
is about 0.5 where we defined $\lambda=W/W'-1$.
When $E_{\rm F}=-0.250$ eV,
$W'$ is about 0.27 eV and $\lambda \approx 0.1$.
We observe that 
the value of $\Gamma_{E_{\rm F}}(k=\pi)$ 
is about 0.13 eV when $E_{\rm F}=-0.054$ eV.
The calculated 
$\epsilon_{E_{\rm F}}(k)$ and 
${\rm Im}(\Sigma(k,\epsilon_{E_{\rm F}}(k)))$
strongly depend on the value of $E_{\rm F}$.

The values of $W'$ and $\Gamma_{E_{\rm F}}(k)$
relate to the peak position and width of $D(V,R^t)$.
We plot $D(V,R^t)$ for $E_{\rm F}=-0.054$ eV 
at $R^t=0.42$, $0.84$, $1.26$ nm
in Fig.~\ref{fig:selfene}(b).
The LDOS curves have several sharp spikes 
at $V=\epsilon_{E_{\rm F}}(k)$ due to 
relatively small value of $\Gamma_{E_{\rm F}}(k)$
compared with the finite level spacing of the edge states.
Here we take $(120,0)$ CNT ($d_t\approx 10$ nm)
for calculating the self-energy.
For a graphene ($d_t \to \infty$),
the level spacing becomes zero and 
the spike structure will disappear.
The calculated LDOS structure for $E_{\rm F}=-0.054$ eV 
has a peak near $V \approx -60$ meV 
which is close to the observed LDOS in which 
a peak is located at $V = -30\sim -20$ 
meV.~\cite{niimi05,niimi06,kobayashi05,kobayashi06}
However, the width of the peak is about 0.7 eV,
which is much larger than the experiment~\cite{niimi05,kobayashi05}
($0.05 \sim 0.1$ eV).
The peak position and the width are improved to fit the experimental
data when $W=0.1$ eV as shown in Fig.~\ref{fig:selfene}(c) and (d).
In this case, 
$W'=0.02$ and $0.04$ for $E_{\rm F}=-0.002$ and $-0.27$ eV, 
respectively.
The calculated LDOS structure for $E_{\rm F}=-0.027$ eV 
has a peak near $V \approx -20$ meV 
with the width of $\approx 0.05$ eV.
The small value of $W$ does not show any spike structure 
due to the discrete $k$ values.
Thus overall feature seems to be better for $W = 0.1$ eV than 
$W=0.3$ eV.
It should be noted that
$W=0.1$ eV is not always consistent to the experiment.
When we assume all the edge states are below the Fermi energy,
we see that $E_{\rm F}$ appears above $E=0$ eV which gives a peak 
very close or above the $V=0$ eV, while the experiment shows 
$V = -30\sim -20$ meV.

It is worth mentioning that
Niimi {\it et al}.~\cite{niimi05,niimi06}
reported that the tunnel current was unstable
when the STS tip was located very close to the zigzag edge.
Then the electron injected from the STS tip has 
a large transition amplitude 
to edge states having $\xi(k) < 2\ell$ (0.42 nm), 
that is, $k$-states which are 
around $k=\pi$ state (see Fig.~\ref{fig:edgestate}(d)).
As shown in Fig.~\ref{fig:selfene}(c),
the magnitude of 
$\Gamma_{E_{\rm F}}(k)$ is much larger than 
$k_{\rm B} T$ ($\approx$ 0.0045eV)
for most value of $k$ and
$\Gamma_{E_{\rm F}}(k)$ for states around $k=\pi$ state
reaches about 0.02 eV and yield strong fluctuation.
It indicates that the tunnel current is unstable.
We calculate $D(V,0)$ and find that 
the height and width of the peak are both significantly larger
(more than 10 times larger) than $D(V,R^t)$ for $R^t=0.42$ nm.
As we noted in Sec.~\ref{sec:ed-ph},
iTO and LO modes give a large matrix element 
through the boundary deformation potential.
The boundary deformation potential
may be relevant to the unstable tunnel current.
Since the injected electron from a STS tip
is localized near the edge,
we expect that the tunnel current would be strongly 
affected by the boundary deformation potential.

\section{discussion and summary}\label{sec:dis}

We showed that el-ph interaction for edge states
consists only of the on-site atomic deformation potential.
As a result, 
LA mode contributes to scattering most effectively and
the on-site deformation potential 
is enhanced at the edge for LO and iTO modes.
It is to be noted that
the on-site atomic deformation potential 
does not contribute to backward scattering of extended states
and the off-site atomic deformation potential
gives rise to resistivity.~\cite{suzuura02}
Because the on-site atomic deformation potential 
is larger than the off-site atomic deformation potential, 
the edge states exhibit the strong el-ph coupling character 
that the graphite system originally possesses.

The original energy bandwidth of the edge states, $W$, is 
consistent with the observed position of LDOS peak 
when $W\approx 0.1$ eV, which is 
the same order of $\gamma_n \approx 0.3$ eV but not the same value.
It is noted that the case of $W = 0.3$ eV 
does not include the overlapping integral
($s$ parameter~\cite{saito98book}) 
which increases (decreases) the conduction (valence) bandwidth.
To examine the effect of $s$ parameter on 
the bandwidth of the edge states, we performed 
the energy band structure calculation in an extended tight-binding
framework~\cite{samsonidze04} and obtained $W \approx 0.2$ eV.
We expect that $W$ 
is externally modified by attaching a functional group
or contacting an electrode to the edge sites.
In addition to the el-ph interaction, 
electron-electron (el-el) interaction causes self-energy which 
may account for the difference.
The el-el interaction contributes to $\Sigma(k,i\omega_n)$
additively when there is no cross term of el-ph and el-el interactions.
Since many papers report the importance of el-el interaction,
especially the contribution to the imaginary part of 
self-energy should be important and thus the spike structure
of discrete $k$ does not appear in the real case.
Then $W'$ decreases and the width of LDOS increases.
The effect of el-el interaction 
warrants future work concerning the effects of 
dynamical details on LDOS curve.
A detail experimental data of STM/STS 
for zigzag edge of CNTs and graphene
may be useful for a qualitative estimation of the strength of 
the el-el interaction.
When the values of $W$ and $\hbar \omega_{\rm D}$ are comparable,
the vertex correction may be important since the Migdal theorem
is not applicable.
Although our results are consistent with the STS data,
it is possible that vertex correction may change 
$\alpha_{kk'}^\nu({\bf q})$.
However the vertex correction to $\alpha^\nu_{kk'}({\bf q})$
is beyond our scope in the present paper.

We did not consider the el-ph matrix element 
between extended state and edge state.
Although the matrix element may be enhanced 
by the boundary deformation potential, 
it is naively expected that 
the matrix element is proportional to 
$\sqrt{\xi/L}$ and is negligible when $L \gg \xi$.
In this case,
the extended state and the edge state can be decoupled.
The geometry with $d_t \gg L$ is referred to as 
the nano-graphite ribbon. 
For a ribbon, the off-site atomic deformation potential 
contributes to the scattering between two edge states 
which are located at the different edges
since overlapping between the two edge states
is not negligible.
It is noted that Igami {\it et al}.~\cite{igami98}
showed that out-of-plane edge phonon modes appear 
depending on the effective mass of carbon atom at edge sites,
and 
Tanaka {\it et al}.~\cite{tanaka02} observed such modes 
at the armchair edge of nano-graphite ribbons on TiC(755) surface
by high resolution electron energy loss spectroscopy.
Though ${\bf e}^\nu_{\bf q}(s)$ and $\omega_\nu({\bf q})$
used in this paper do not include the edge phonon mode,
el-ph coupling for out-of-plane modes 
is negligible for the edge states.
The el-ph interaction for nano-ribbon requires 
further studies on the el-ph interaction.

In summary,
we formulated el-ph interaction for edge states
and used it to calculate LDOS.
Although our calculation does not include the Coulomb interaction,
the result agrees with LDOS data~\cite{niimi05,niimi06,kobayashi05} 
when $W\approx 0.1$ eV.
Our results should be compared with the future experiments
of edge states in CNTs and graphene.

\begin{acknowledgments}
 R. S. acknowledges a Grant-in-Aid (No. 16076201) from MEXT.
 K. S. acknowledges G. Samsonidze (MIT) for using programs 
 calculating phonon dispersion relation and eigenvector.
\end{acknowledgments}


\end{document}